\documentclass[%
 reprint,
superscriptaddress,
 amsmath,amssymb,
 aps,
]{revtex4-2}

\usepackage{graphicx}% Include figure files

\usepackage{dcolumn}% Align table columns on decimal point
\usepackage{bm}% bold math
\usepackage[dvipsnames]{xcolor}

\usepackage{todonotes}

\begin{document}

\preprint{APS/123-QED}

%\title{Phase synchronization of oscillators from parametric resonance near ghost attractors}% Force line breaks
%\title{Nematic order from phase synchronization of deformable tumbling particles\\\su{Nematic order from phase synchronization of shape oscillations}}% Force line breaks
\title{Nematic order from phase synchronization of shape oscillations}% Force line breaks
%\title{Dynamic nematic order via phase synchronization in a dilute suspension of tumbling deformable particles}% Force line breaks

\author{Ioannis Hadjifrangiskou} 
\affiliation{Rudolf Peierls Centre for Theoretical Physics, University of Oxford,  Oxford OX1 3PU, United Kingdom}
%\email{ioannis.hadjifrangiskou@physics.ox.ac.uk}
\author{Sumesh P. Thampi} 
\affiliation{Rudolf Peierls Centre for Theoretical Physics, University of Oxford,  Oxford OX1 3PU, United Kingdom}
\affiliation{Department of Chemical Engineering, Indian Institute of Technology Madras, Chennai-36, India}
%\email{ioannis.hadjifrangiskou@physics.ox.ac.uk}
%\author{Julia M. Yeomans$^{1}$} %\email{ioannis.hadjifrangiskou@physics.ox.ac.uk}
\author{Rahil N. Valani}\email{rahil.valani@physics.ox.ac.uk}
\affiliation{Rudolf Peierls Centre for Theoretical Physics, University of Oxford,  Oxford OX1 3PU, United Kingdom}
%\affiliation{$^2$School of Physics and Astronomy, Monash University, Victoria 3800, Australia}

\date{\today}% It is always \today, today,

\begin{abstract}
% A deformable elongated particle undergoes rotation and shape oscillations in a steady shear flow. We show that such a particle suspended in a time-dependent shear flow, with a dominant steady shear and a small amplitude oscillatory component, can synchronize it tumbling motion and shape oscillations to driven shear oscillations. For a dilute suspension non-interacting deformable particles, this leads to the development of nematic order arising from phase synchronization. Our results also extend to a dense suspension of elongated particles where the equations of spatially averaged shape deformations and orientations also lead to a nematic phase from phase synchronization as opposed to inter-particle interactions. This emergence of nematic order in driven shear flows can also be realised in steady shear flow with an active source of oscillations, either as active shape oscillations or pulsating chiral activity. Our results motivate experiments of driven shear flow with deformable elongated particles where such phase-synchronized nematic order might be observable and affect rheological properties of soft matter and active systems.

% A deformable particle undergoes rotation and shape oscillations in a shear flow. 
We show that a suspension of non-interacting deformable particles subjected to an oscillatory shear flow leads to development of nematic order that arises from the phenomenon of phase synchronization. The synchronized state corresponds to a unique, stable limit cycle confined in the toroidal state space. The limit cycle exists since, unlike rigid particles, deformable particles can modulate aspect ratio, adjust their tumbling rate and thus, achieve phase synchronization. These synchronized regions emerge as Arnold tongues in the parameter-space of the driving amplitude and frequency. Considering the rheological implications of ordering dynamics in soft and active matter, our results motivate oscillatory shear flow experiments with deformable particles.
% Our results also extend to a dense suspension of elongated particles where the equations of spatially averaged shape deformations and orientations also lead to a nematic phase from phase synchronization as opposed to inter-particle interactions. This emergence of nematic order in driven shear flows can also be realised in steady shear flow with an active source of oscillations, either as active shape oscillations or pulsating chiral activity. 

%then although they tumbling at the same frequency, their orientation i.e. the phase, will be different for different particle. We show a mechanism of synchronizing the phases of these tumbling particles by driving the system with oscillating shear flow in addition to constant shear. We find resonance tongues and synchronization regimes where the deformation and tumbling motion of the particles get phase and frequency locked onto the driven shear oscillation. At the collective level, this results in an emergent dynamic nematic order of the system due to synchronization from non-equilibrium driving.

\end{abstract}

\maketitle

% \textit{Introduction--} 
Unlike solids and simple liquids, orientational order of constituent particles is a defining feature of soft and biological matter \cite{van2024soft, ma2022self, needleman2017active, shankar2022topological}. Classic theories predict the emergence of nematic order: Maier-Saupe theory based on anisotropic attractive interactions \cite{maier1958einfache, stephen1974physics} and Onsager's theory based on the gain in translational entropy at the expense of lowered orientational entropy of the particles \cite{onsager1949effects}. Athermal, active and external fields \cite{lelidis1993electric, olmsted1990theory, ostapenko2008magnetic, pearce2019activity, mueller2019emergence} can also drive the isotropic - nematic transition. Notably, all these approaches rely on interaction between anisotropic particles. They also ignore a characteristic feature ubiquitous in soft and biological matter, namely the deformability of the constituents e.g., as in polymers, soft colloids, emulsions and cells \cite{barrat2023soft,manning2023essay}. In this {letter}, we show how shape deformability in conjunction with shape anisotropy can induce nematic order of non-interacting particles in a periodically driven system, via a phase synchronization mechanism.

It is well known that two weakly coupled oscillators can synchronize their oscillations~\citep{Pikovsky_Rosenblum_Kurths_2001}. Examples of synchronization are ubiquitous, such as flashing fireflies \citep{buck1988synchronous}, pendulum clocks \citep{rosenblum2003synchronization}, mammalian cell cycles \citep{davis2001biological}, beating cilia of microorganisms \citep{uchida2017synchronization,Golestanian2011}, and optomechanical and nanomechanical oscillators~\citep{Georg2011,Holmes2012}. A particular case of synchronization is the entrainment of an oscillator by an external, periodic driving. In this case, the coupling is unidirectional, i.e. the external driving can influence the oscillator but not vice versa. Then the frequency and phase of the driven oscillator gets locked onto the driving signal; and this is known as \emph{phase synchronization}~\citep{Pikovsky_Rosenblum_Kurths_2001,Pikovsky2000}. Examples of systems that exhibit phase synchronization due to entrainment are hair cells of the inner ear~\citep{Hemsing2012,Levy2016}, Josephson junctions~\citep{Likharev2022-yn,Gandhi2015} and driven colloidal systems in periodic potentials~\citep{Juniper2015,Juniper_2017}. 

In this work, we consider a system of deformable, tumbling particles dispersed in a simple shear flow as independent oscillators and drive them with a time-dependent, periodic shear. We show that the orientational and shape oscillations of the particles phase synchronize to the oscillatory shear driving. For many such non-interacting deformable particles, this phase synchronization leads to the emergence of nematic order. As opposed to particle-particle interactions that are traditionally responsible for nematic order in liquid crystals ~\citep{de1993physics,beris1994thermodynamics} or synchronization in Kuramoto models~\citep{kuramoto1975self}, the proposed mechanism is rooted in non-equilibrium driving and is not applicable for rigid particles.%Characteristic of soft matter systems, the concerned particles are shape deformable.

% show that for deformable elongated particles in a time-dependent shear flow, where a small amplitude oscillatory component is added to a steady shear flow, the orientational and shape oscillations of the particle phase synchronize to this oscillatory shear driving. For many such non-interacting deformable particles, this phase synchronization leads to the emergence of nematic order that is rooted in non-equilibrium driving as opposed to particle-particle interactions that are traditionally responsible for nematic order in liquid crystals or synchronization in Kuramoto models~\citep{de1993physics,beris1994thermodynamics, kuramoto1975self}.% \io{Can make this a more general statement I think - thermodynamic interactions are used even outside the context of LCs to align particles (e.g. Kuramoto-like terms)}
%. 
%while synchronization from external driving has been discussed in various aspects of physics and biology, it has not been investigated much in soft matter research.

\begin{figure}
\centering
\includegraphics[width=\columnwidth]{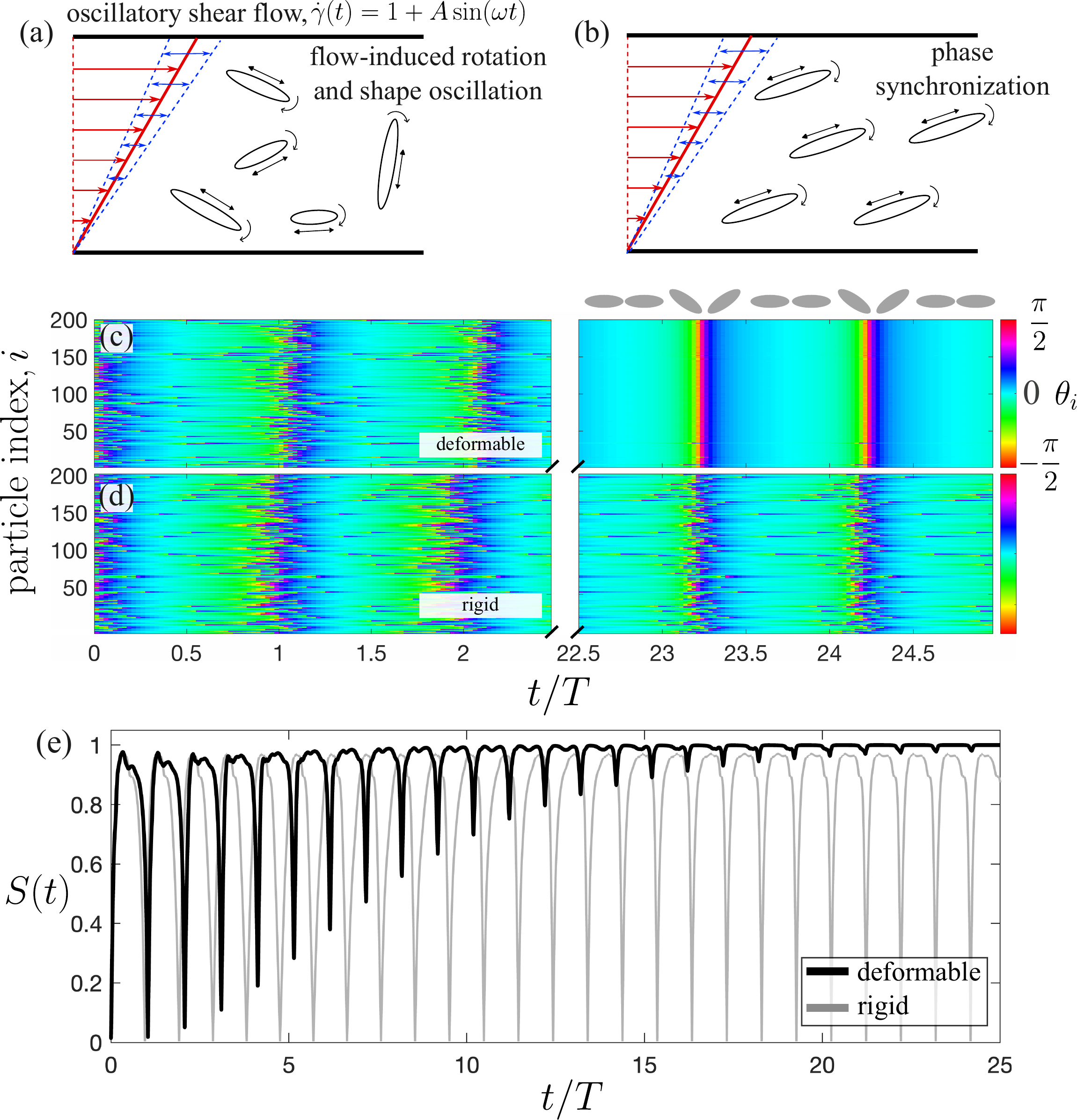}
\caption{Emergence of nematic order for non-interacting deformable particles in an oscillatory shear flow. %, $\dot{\gamma}(t)=1+A \sin(\omega t)$ with $T=2\pi/\omega$. 
(a) Schematic of the system showing deformable particles suspended in a time-dependent shear flow with both a mean (red) and an oscillatory (blue) component. The initial isotropic state of random orientations $\theta_i(0)$ and varying shape parameters $r_i(0)$ develops into (b) a phase synchronized state at later times.  Early and late time kymographs illustrating the temporal evolution of orientation of (c) deformable and (d) rigid particles. (e) The corresponding nematic order parameter $S(t)$ for deformable (black) and rigid (grey) particles. Grey ellipses at top of (c) are a guide for particle orientation while phase synchronized. The plots are for $N=200$, $A=0.5$ and $T=20$.}
%The top and bottom plots of (c) show the evolution of orientation $\theta_i$ for each particle during initial times for deformable and rigid particles, respectively. (b) Schematic showing a phase-synchronized state and (d) the corresponding evolution of $\theta$ at later times for deformable particles (top) clearly showing phase synchronization, whereas rigid particles don't synchronize (bottom). 
%Grey circles at top of (d) are a guide for particle orientation during phase synchronization of deformable particles. (e) Evolution of nematic order parameter $S(t)$ for (black) deformable and (grey) rigid particles showing the emergence of nematic order for deformable particles due to phase synchronization. The parameters are fixed to $A=0.5$, $T=20$ and $N=200$.}
%In the synchronized state, each particle's motion is entrained by the oscillatory driving where (d) the phase difference $\theta(t)-\phi$ stays bounded and locked into oscillatory evolution and (e) the period of tumbling motion adjusts from its natural period $T_n$ to the driving period $T$.}
%\su{Fig 1: Replace r Kymos with that of rigid particles to show the difference? Add S of rigid particles in (c) ?}
\label{Fig 1}
\end{figure}

% \begin{figure}
% \centering
% \includegraphics[width=\columnwidth]{Fig2_draft.pdf}
% \caption{Phase synchronization of a single deformable particle to time-dependent shear flow. Each particle's motion is entrained by the oscillatory driving where (top) the phase difference $\theta(t)-\phi$ between the tumbling motion and the external driving $\phi=\omega t$ stays bounded, and (bottom) the period of tumbling motion modulates to the driving period.}
% \label{Fig 2}
% \end{figure}

\begin{figure*}
\centering
\includegraphics[width=1.8\columnwidth]{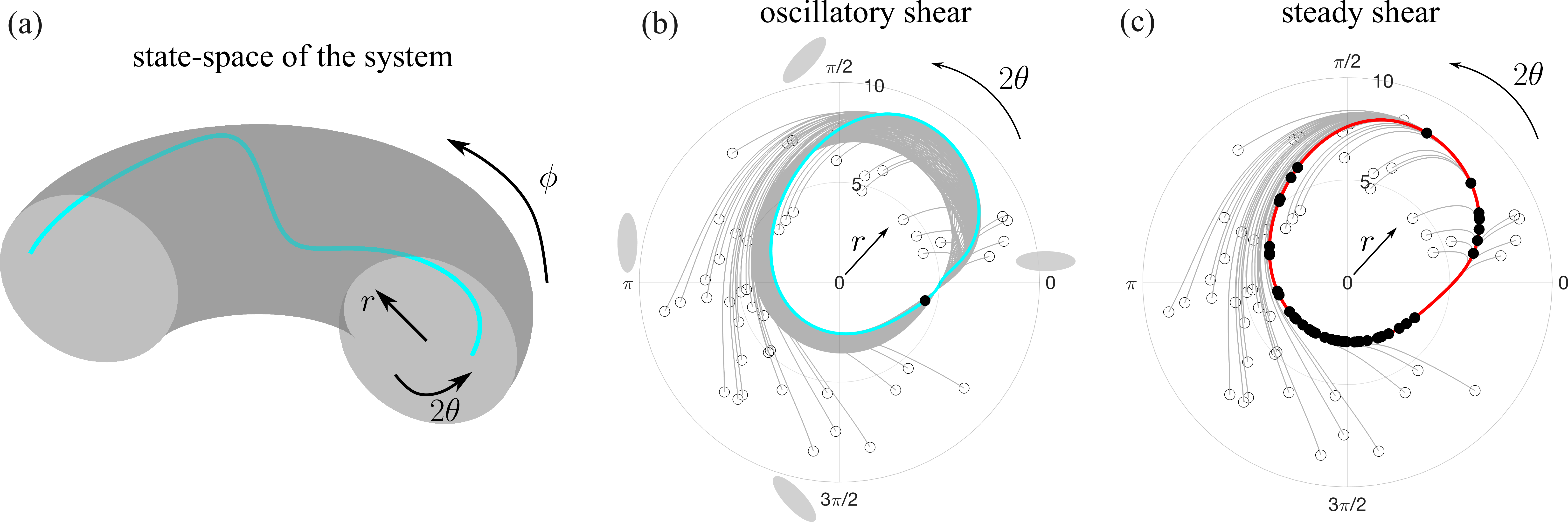}
\caption{State space description. (a) Schematic of the $3$D state-space represented by a torus -  driving phase $\phi$ as the toroidal angle, particle orientation $2\theta$ as the poloidal angle and particle shape parameter $r$ as the smaller radius. Projection of the torus
%the $3$D state space %of the limit cycle of this driven system with oscillatory shear%(i.e. $\dot{\gamma}=1+A\sin(\omega t$)) 
onto the $(r,2\theta)$ polar plane for a system subjected to (b) oscillatory shear ($A \neq 0$) and (c) steady shear ($A=0$). %co-ordinates showing that the same 
Random initial conditions (open circles) converge to a point (filled circle) on the limit cycle (cyan) in (b), 
%.  (c) Polar $(r, 2\theta)$ state space of the non-oscillatory system($A=0$). %(i.e. $\dot{\gamma}=1$) in the polar phase-space formed by the shape parameter $r$ (radial co-ordinate) and orientation $\theta$ (angular co-ordinate is $2\theta$). 
%in the $(r, 2\theta)$ polar state space. 
% Random initial points 
but they converge at different phases on the limit cycle (red) in (c).} %(% $(x,y,z)=((R+r\cos(2\theta))\cos(\phi),(R+r\cos(2\theta))\sin(\phi),r\sin(2\theta))$. 
\label{Fig 2}
\end{figure*}

\begin{figure}
\centering
\includegraphics[width=0.75\columnwidth]{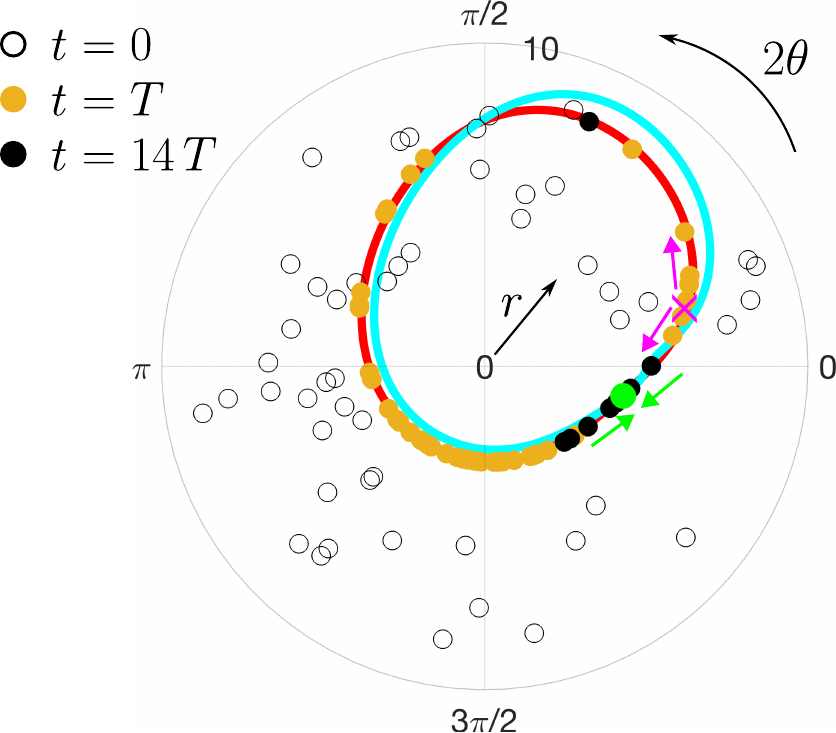}
\caption{Transient dynamics in state-space. Snapshots of trajectories in the state-space as points while passing through the $(r, 2\theta)$ cross section of the torus at three different times. At $t = 0$, random initial conditions (open circles); at $t = T$, trajectories converged to the limit cycle of the steady shear system (yellow filled circles on the red curve); and at $t = 14T$, trajectories (black filled circles)  converging onto the stable point (green filled circle with incoming arrows) that lies on the limit cycle of the oscillatory system (cyan curve). The unstable point (pink cross with outgoing arrows) corresponds to the unstable limit cycle in the full state-space.}
\label{Fig 2_2}
\end{figure}

\begin{figure*}
\centering
\includegraphics[width=2\columnwidth]{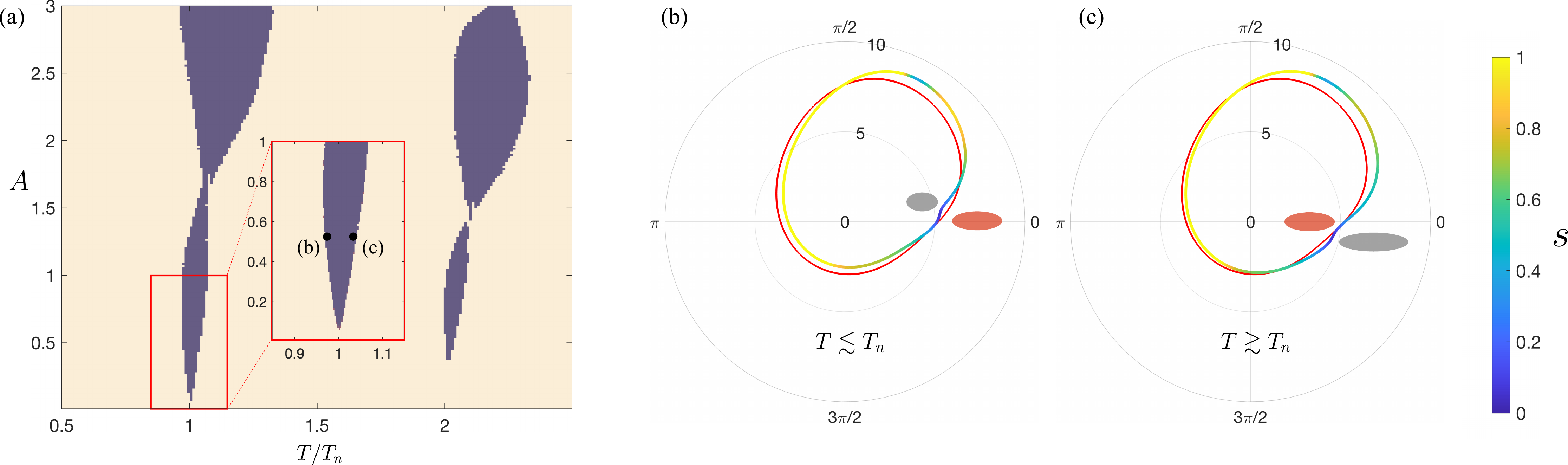}
\caption{(a) Arnold tongues: regions of phase synchronization induced nematic order (violet) in the $(A,T)$ parameter space. The inset shows the expanded view of the first tongue. %formed by the amplitude $A$ and period $T$ of shear oscillations. 
% Arnold tongues %that are characteristic of synchronization, 
% are observed for increasing $A$ at multiples of the natural period of the limit cycle $T_n$. 
%(b)-(d) Poincare sections showing the saddle-node bifurcation on the undriven limit cycle (red) for $T = 19.5$, these points spread on the limit cycle near the middle of the tongue ($T=20$) (c) and then annihilate again in a saddle-node bifurcation at the other end of the tongue ($T=20.78$).
(b)-(c) Limit cycles in $(r,2\theta)$ projection of the state-space, corresponding to two points marked in (a): $T=19.5\lesssim T_n$ in (b) and $T=20.7\gtrsim T_{n}$ in (c). The color on the limit cycle denotes the state-space speed $s=\sqrt{(dr/dt)^2+(d\theta/dt)^2}$. In each figure, the limit cycle for the case of steady shear ($T_n\approx 20.02$) is plotted in red. The grey and red ellipses illustrate the aspect ratio of the deformed particle (not to scale) as they pass through the ghost region (near the orientation $\theta=0$) in the oscillatory and in the steady shear flow respectively. Compared to particles in steady shear, in (b) particles reduce their anisotropy as they pass through the ghost region to reduce their period. Conversely, in (c) particles increase their anisotropy to increase their period.%in (b) the particle reduce its   and the grey ellipses illustrate the shape changes (not to scale). In the former (latter) figure particles reduce (increase) their anisotropy as they pass through the ghost region ($\theta=0$) to reduce (increase) their period. 
}
% driving period is less than the limit cycle (red) period ($T=19.5<T_n$ with $T_n\approx 20.02$) and (c) provide physical insight into shape deformations in synchronizated state. In the $(r,2\theta)$ projection of the state-space, trajectories spent most of their time near the ghost region around $\theta\approx0$. (b) When the driving period is less than the limit cycle (red) period ($T=19.5<T_n$ with $T_n\approx 20.02$), particles reduce their shape parameter $r$ as they pass through the ghost region (color indicates speed) to reduce their period. (c) Conversely, when the driving period is larger than the limit cycle period ($T=20.7>T_{n}$), particles increase their shape parameter as they pass the ghost region to increase their period and match with the driving period.

%ing that for a fixed $A$ as one varies $T$, these synchronized regions come into existence via a saddle-node bifurcation on the undriven limit cycle (red) $T=19.5$, these points spread on the limit cycle near the middle of the tongue ($T=20$) (c) and then annihilate again in a saddle-node bifurcation at the other end of the tongue ($T=20.78$).}
\label{Fig 3}
\end{figure*}

% \begin{figure}
% \centering
% \includegraphics[width=\columnwidth]{Fig4_NEW_V3.pdf}
% \caption{Physical insight into shape deformations that lead to synchronization. In the $r-2\theta$ projection of the phase-space, trajectories spent most of their time near the ghost region around $\theta\approx0$. (a) When the driving period is less than the limit cycle (red) period ($T=19.5<T_0$ with $T_0\approx 20.02$), particles reduce their shape parameter $r$ as they pass through the ghost region (color indicates speed) to reduce their period. (b) Conversely, when the driving period is larger than the limit cycle period ($T=20.7>T_0$), particles increase their shape parameter as they pass the ghost region to increase their period and match with the driving period.}
% \label{Fig 4}
% \end{figure}

\textit{Model--} Consider a system of $N$ deformable particles subjected to a time-dependent shear flow $\dot{\gamma}_0(1+A \sin(\omega t))$, as schematically shown in Fig.~\ref{Fig 1}(a). The instantaneous state of the $i^{th}$  particle, in two dimensions, is described by two degrees of freedom, (i) a shape parameter $r_i$ equal to the aspect ratio of the particle minus unity, and (ii) an inclination angle $\theta_i$ indicating the axis of elongation measured with respect to the flow direction. $r_i=0$ corresponds to a circular particle whereas larger $r_i$ corresponds to a more elongated particle. The dynamics of the $i^{th}$ particle are described by
\begin{align}%\label{eq: full}
 \frac{\text{d} {r}_i}{\text{d} t}&= \dot{\gamma}(t) (r_i+1)\sin{2\theta_i} - \frac{1}{\alpha}(r_i-r_{0})\left(1 +  \epsilon (r_i-r_{0})^{2}\right), \label{eq:reqn}\\
  \frac{\text{d} {\theta_i}}{\text{d} t}&= \frac{\dot{\gamma}(t)}{2}\left(\beta(r_i) \cos{2\theta_i} - 1\right). \label{eq:thetaeqn}
\end{align}
In Eqs.~\eqref{eq:reqn}-\eqref{eq:thetaeqn},  $\dot{\gamma}(t)=1+A \sin(\omega t)$, $\beta(r_i)=\left((r_i + 1)^2 - 1\right)/\left((r_i+1)^2 +1\right)$ and time is scaled with the inverse of the mean shear rate $\dot{\gamma}_0^{-1}$.

Eq.~\eqref{eq:reqn} describes the evolution of the shape parameter~\citep{bilby1977finite,gao2011rheology, hadjifrangiskou2023active}.  The first term on the right-hand-side captures the strain-induced deformation of the particle due to velocity gradients in the imposed flow, while the second term is derived from a free energy density, $f_{r_i} =A_{r} [\frac{1}{2}(r_i-r_{0})^2  +  \frac{\epsilon}{4} (r_i-r_{0})^{4}]$ that maintains an equilibrium particle shape with shape parameter $r_{0}$ in the absence of flow. Here $\alpha=\dot{\gamma}_0/(\Gamma_{r}A_{r})$ controls the strength of free energy induced shape change compared to that of the shear,  $\epsilon$  describes the quartic free energy landscape and the parameter $\Gamma_{r}$ is a relaxation rate towards the free energy minimum at $r_i = r_{0}$.

Eq.~\eqref{eq:thetaeqn} describes the angular velocity of the deformed, anisotropic particle through Jeffery's equation~\citep{jeffery1922motion}. The two terms in the right-hand-side parenthesis respectively capture the effect of the strain rate and vorticity components of the shear flow. Therefore, the angular frequency of the tumbling motion depends on the instantaneous shear rate $\dot{\gamma}(t)$, shape parameter $r_i(t)$ and orientation $\theta_i(t)$. The rotational velocity induced by the strain rate (first term) is orientation dependent, and hence, even rigid anisotropic particles ($r_i=$ constant) rotate at a non-uniform rate in a steady shear flow ($A = 0$).

We numerically solve Eqs.~\eqref{eq:reqn}-\eqref{eq:thetaeqn} for $N= 200$ particles with random initial orientations (isotropic state). Without loss of generality, we fix $\alpha=0.8$, $\epsilon=0.1$ and $r_0=5$ in the following analysis. 

\textit{Emergence of nematic order --} Simulations show that, over time, all particles tumble in the flow at the same driving frequency $\omega$, thus frequency locking their rotational motion to the imposed oscillatory shear flow. More importantly, we find that the particles also phase lock their rotational motion with the imposed flow~(see Sec.~IB of \citep{supplementary_m}). This phase locking of each particle's orientation with the imposed oscillatory flow implies the synchronization in the motion of all particles with each other as well, and we obtain a state of collective behavior where orientations of all non-interacting deformable particles are the same at all, sufficiently long, times ($t \gg{2\pi}/{\omega}$). Thus, nematic order emerges in the system as indicated schematically in Fig.~\ref{Fig 1}(b) (see Supplemental Video S1). 
% In the absence of periodic driving ($A=0$), namely in a steady simple shear flow, all the particles tumble with a time period $T_n = \int_{0}^{\pi}(d\theta_i/dt)^{-1} d\theta$, referred to as the natural time period of the system. Thus individual particles are frequency locked with the flow, with a time period dictated by shear. However, the phase of their tumbling is determined by the initial conditions. Hence, collectively the system continues to be in isotropic state. On the other hand, driving the system periodically ($A \neq 0$) with a period $T \neq T_n$, the natural time period $T_n$ becomes irrelevant.  

Kymographs illustrating the evolution of the orientation of particles $\theta_i(t)$ are shown in Fig.~\ref{Fig 1}(c) for early and late times. The isotropy in the initial orientation of particles gradually fades away with time, and at long times the phase-synchronized state emerges with tumbling frequency $\omega$ and no variations in $\theta_i$ between different particles. 
%The phase-synchronized state continues to exhibit periodic oscillations at the driving frequency $\omega$. 
In contrast, the kymograph obtained for rigid, elongated particles, shown in Fig.~\ref{Fig 1}(d) does not exhibit phase locking and the system does not develop nematic order~(see Supplemental Video S2). 

To quantify the nematic order in the system, we calculate the magnitude of the nematic order parameter $S(t)=\langle\cos\left[2(\theta_i(t)-\Psi_0(t))\right]\rangle $ %~\citep{add reference for this equation} no need for reference here I think. I will add the derivation of getting \Psi_{0} in the supplementary and we can decide whether we actually want to include it.
where $\langle\cdot \rangle$ denotes an average over $N$ particles and $\Psi_0(t)$
% $=\dfrac{1}{2}\tan^{-1}\left( {\sum_{i}^N \sin(2\theta_i(t))}/{\sum_{i}^N \cos(2\theta_i(t))} \right)$ 
is the mean orientation at time $t$. The evolution of $S(t)$ for deformable particles is shown in Fig.~\ref{Fig 1}(e) (black curve) which asymptotes towards unity indicating the development of permanent nematic order in the system. The slow development of nematic order is due to the difference in time taken to phase synchronize based on initial conditions~(see Sec.~ID of \citep{supplementary_m}). Conversely, the measured nematic order parameter in a system of rigid particles subjected to the same oscillatory flow (gray curve in Fig.~\ref{Fig 1}(e)) keeps oscillating between $0$ and $\lesssim 1$~\footnote{Nevertheless, it is still interesting to note that even rigid elongated particles periodically develop nematic order ($S\approx 0.9$) due to critical slowing down of tumbling motion when aligned with the flow i.e. near $\theta_i=0$.}. This analysis reveals that phase synchronization, and hence the emergence of nematic order, is a direct consequence of the deformability of particles. 

%in a system of rigid particles subjected to the same oscillatory flow keeps fluctuating with $S(t)$ dropping to zero periodically (gray curve in Fig.~\ref{Fig 1}(e)).

\textit{Phase synchronization --} We now use a general mathematical framework~\citep{Pikovsky_Rosenblum_Kurths_2001} to demonstrate phase synchronization arising from particle deformability. Representing the dynamical system, Eqs.~\eqref{eq:reqn}-\eqref{eq:thetaeqn}, in vector form,
\begin{equation*}
    \dot{\mathbf{x}}_i=F(\mathbf{x}_i;\dot{\gamma}),
\end{equation*}
where $\mathbf{x}_i=(r_i(t),\theta_i(t))$, $F(\cdot)$ is the right-hand-side function, and $\dot{\gamma}(t)=1+A \sin(\omega t)$ is the parametric driving. For $A\ll 1$, doing a perturbation expansion up to linear order in $A$ we obtain
\begin{equation*}
\dot{\mathbf{x}}_i\approx F(\mathbf{x}_i)|_{\dot{\gamma}=1}+A\,\frac{\partial F(\mathbf{x}_i)}{\partial \dot{\gamma}}\Bigg{|}_{\dot{\gamma}=1} \sin(\omega t).
\end{equation*}
Performing a phase reduction~\citep{Pikovsky_Rosenblum_Kurths_2001} from the state variables $(r,\theta)$ to a phase variable $\psi$ that varies along the oscillation (see Sec.~II of \citep{supplementary_m}), and integrating the resulting equation over one period of the parametric driving $T=2\pi/\omega$, we obtain the circle map
\begin{equation}\label{eq: circle map gen}
\psi_{k+1}=\psi_{k}+2\pi \frac{\omega_n}{\omega} + A\, f(\psi_k),
\end{equation}
where $\omega_n$ is the natural frequency  (\textit{i.e.,} for $A=0$) and $f(\psi_k)=\int_{k T}^{(k+1)T} \frac{\partial \psi}{\partial r_i} \frac{1}{\alpha}(r_i-r_0) \left(1 +  \epsilon (r_i-r_{0})^{2}\right) \sin(\omega t) dt$. The map represented by Eq.~\eqref{eq: circle map gen} exhibits phase synchronization in general~\citep{Jackson_1989}. On the other hand, for rigid particles $f(\psi_k)=0$ and they do not phase synchronize. We now turn to investigate the underlying dynamical mechanism responsible for \emph{phase synchronization} of the deformable, tumbling particles.

\textit{Dynamical analysis--}
 Eqs.~\eqref{eq:reqn}-\eqref{eq:thetaeqn} constitute a non-autonomous dynamical system due to the time-dependent driving $\dot{\gamma}(t)$, but introducing a new phase variable  $\phi_i=\phi=\omega t$ or,
\begin{align}
   \frac{\text{d} {\phi}}{\text{d} t} &= \omega, \label{eq: full}
\end{align} 
 with $\phi(t+2\pi/\omega)=\phi(t)$, results in Eqs.~\eqref{eq:reqn}, \eqref{eq:thetaeqn} and \eqref{eq: full} to form an autonomous $3$D dynamical system for a single particle  ~(see Sec.~IA of \citep{supplementary_m} for more details) and allow us to pursue a state-space analysis. 

A subset of the complete $3$D state-space formed by variables $r,\theta,\phi$
% ($\equiv r_i, \theta_i, \phi_i$) 
for a single deformable particle can be represented as the inside of a torus with $\phi$ the toroidal angle, $2\theta$ the poloidal angle and $r$ the smaller radius of the torus ~(see Fig.~\ref{Fig 2}(a)). In the synchronized state we find that $N$ different initial conditions corresponding to $N$ particles converge to a stable limit cycle orbit that lies inside the torus (the cyan curve). Furthermore, along the limit cycle all $N$ trajectories converge to the same point at any given time. This is better illustrated in Fig.~\ref{Fig 2}(b) where the azimuthal projection of the torus on to a polar plane formed by $(r,2\theta)$ is shown. The $N$ trajectories not only converge to the projected limit cycle (cyan curve) but also to the same point  (black dot) at a given time indicating phase synchronization.

The focusing of state-space trajectories onto a point on the limit cycle can be understood as follows. All trajectories in state-space rotate at the same angular velocity $\frac{\text{d} {\phi}}{\text{d} t} = \omega$ around the torus. This implies that $N$ random initial conditions which start at the same initial time (say, $\phi(0)=0$) must occupy the same instantaneous cross-section of the torus -- the polar plane formed by $r$ and $2\theta$, at any future time $\phi(t)$. Since the state-space trajectories must eventually converge onto the stable limit cycle that wraps around the torus, there is only one point along the limit cycle at each cross-section, and hence all trajectories will converge towards this point on the limit cycle in each cross-section. %, making them converge to a point on the limit cycle. 

This process may be contrasted with a system subjected to a steady shear flow, \textit{i.e.,} $A = 0$. 
% The trajectories of $N$ particles in the projected, $(r,2\theta)$ state-space, are shown in Fig.~\ref{Fig 2}(c). 
Again, all particles deform and tumble, the trajectories converge to a limit cycle orbit (red curve in Fig.~\ref{Fig 2}(c)) with  period $T_n$ but, since the oscillatory shear is absent, the deformable particles occupy different points along the limit cycle at any given time ~(black filled circles) i.e. their orientations $\theta_i(t)$ and shape $r_i(t)$ are different. Hence, deformable particles subjected to a steady shear do not phase synchronize or develop nematic order~(see Supplemental Video S3).

%The synchronized state of the system will be a curve that lies in this torus (say, the black curve in Fig.~\ref{Fig 2}(c)). 
%Commented out the footnote for now, it seems to take the footer of two pages to show the entire thing.
%~\footnote{The free energy term in our dynamical system ensures that $r$ is near $r_0$. Hence if we only consider $0<r<r_{max}$ making sure that $\text{min}\{r(t)\}>0$ and  and $r_{max} \gg \text{max}\{r(t)\}$. The bigger radius $R$ of the torus is chose to be $R>r_{max}$ to avoid self-interaction of the torus.}. 
% In the synchronized state, . Therefore, all initial conditions occupy the same point on the limit cycle in the synchronized state (Fig.~\ref{Fig 2}(b)) .

%This can be seen by projecting the trajectories onto the 2D, polar plane as shown in Fig.~\ref{Fig 2}(b). All initial conditions converge onto the same point onto the projected limit cycle in the synchronized state. This is in contrast with a system of steady shear flow ($A = 0$), where  particles do not synchronize. Initial conditions corresponding to $N$ deformable particles converge onto the projected limit cycle but they will occupy different phases of the limit cycle at any given time~(black filled circles in Fig.~\ref{Fig 2}(a)). Therefore, in this case the orientation $\theta_i(t)$ for each particle will be different at a given time $t$ even though they are all tumbling at the same frequency.

To further elucidate the transient dynamics of phase synchronization, we plot trajectories on the $(r,2\theta)$ cross-section of the torus, similar to that on a Poincar\'e section, at three different times, $t = 0$ (black open circles), $t=T$ (yellow filled circles) and $t\gg T$ (black filled circles) in Fig.~\ref{Fig 2_2}. The initial, randomly distributed points, first, rapidly converge onto the limit cycle of the non-oscillatory system (red curve) and, subsequently, migrate along this limit cycle and converge onto a stable point (green point), which lies on the actual, stable limit cycle of the oscillatory system (cyan curve). In other words, at short times the system behaves effectively like a non-oscillatory system and state space trajectories rapidly converge onto a `ghost' orbit. 
%This orbit is the limit cycle of the non-oscillatory system. 
At long times, spanning multiple driving periods, the oscillatory driving leads to phase synchronization. Accompanying the stable limit cycle, the unstable limit cycle appears as an unstable point (pink cross) that repels the state-space trajectories in this plot.

\textit{Physical mechanism--} Figure~\ref{Fig 3}(a) shows the range of amplitude, $A$, and oscillatory driving period, $T$, for which phase synchronization occurs. The synchronization regions (violet) occur periodically in bands near multiples of $T_{n}$, the natural time period of tumbling in a steady shear flow ($A=0$). These structures in Fig.~\ref{Fig 3}(a) are the well-known Arnold tongues~\citep{Pikovsky_Rosenblum_Kurths_2001} and they come in and out of existence in parameter space via saddle-node bifurcation of cycles in state space~(see Sec.~IC of \citep{supplementary_m}). 
%Outside the tongue, the particles exhibit quasiperiodic motion. Unlike, typically observed~\citep{Pikovsky_Rosenblum_Kurths_2001}, here the tongues exhibit a modulated increase and decrease in the width with increase in $A$. 

%\su{may be show in SM?}The widths of the Arnold tongues do not increase in size monotonically with $A$, u  motion in the synchronized region for $A>1$ is undergoing complex motion which has components of both

% In the absence of periodic driving ($A=0$), namely in a steady simple shear flow, all the particles tumble with a time period $T_n = \int_{0}^{\pi}(d\theta_i/dt)^{-1} d\theta$, referred to as the natural time period of the system
The finite width of the tongues can be physically rationalized by analyzing the extent to which a deformable particle can modulate the rate of tumbling and shape changes 
% adjust the natural time period $T_n$ of the undriven system 
to match the driving period $T$ of the oscillatory shear. %We may first define a natural time period for the system $T_n$, the time period of tumbling in the absence of oscillations ($A=0$). %  $T_n = \int_{0}^{\pi}(d\theta_i/dt)^{-1} d\theta$,
For example, within the primary Arnold tongue in Fig.~\ref{Fig 3}(a), the particle decreases (increases) its tumbling period on the limit cycle when $T<T_{n}$ ($T>T_{n}$) by suitable shape deformations. For the tumbling dynamics, since the particle spends most of its time near the $\theta=0$ ghost region~\citep{strogatz} (i.e. region of critical slowdown in state space), the shape modulations near this orientation have the most significant contribution to the tumbling period.
% The modulation in the angular velocity attained via shape deformations 
% of the particle at two points on the edge 
% near the tips 
% of the Arnold tongues are shown in 
% Fig.~\ref{Fig 4}(a) for $T=19.5<T_{n}$ and in Fig.~\ref{Fig 4}(b) $T=20.7>T_{n}$ 
This modulation is illustrated in Fig.~\ref{Fig 3}(b) \& (c) for $T <T_{n}$ and $T>T_{n}$ respectively.
% which correspond to two points marked (b) \& (c) in Fig.~\ref{Fig 3}(a) the conditions that lie close to the edge of the tongue. 
In these figures, the limit cycle is colored to denote the state-space speed $s=\sqrt{(dr/dt)^2+(d\theta/dt)^2}$. The limit cycle for the non-oscillatory system (red curve) is also plotted for comparison.
%The red curve corresponds to the limit cycle of the undriven system. 
We can see in Fig.~\ref{Fig 3}(b) that compared to that of the non-oscillatory system, the oscillatory system reduces its shape parameter $r$ near the orientation $\theta=0$. The reduced elongation of the particle helps it traverse the ghost region faster and thus to reduce the period of the limit cycle.  % It does this, even at the cost of a reduction in $r$ near $\theta = \pi/2$.
% which corresponds to the fastest rotational speed
% , which would cause a slow-down. 
% The speeding up near the ghost region more than compensates. We note that at $\theta=0$, the flow is unstable to couple to shape modulations and the free energy component dominates which relaxes the shape parameter to $r_0=5$. Hence, the particle adjusts its shape immediately before and after $\theta=0$. 
Similarly, the limit cycle in Fig.~\ref{Fig 3}(c) has $r$ increased in order to traverse the ghost region slower and thus increase the period of the limit cycle. Hence, by appropriately modulating the shape near the ghost region, the particle matches the period of the limit cycle with the imposed oscillation period. %The extent of the particle's ability to achieve this is related to the system parameters and is made manifest in the finite width of the Arnold tongues. %and the resultant changes in the angular velocity leads to synchronization of the deformable particles with imposed oscillations.
%\rv{Add a paragraph describing Fig 4 ...}

\textit{Discussion and Outlook --}  
We have shown that deformable anisotropic particles can phase-synchronize in time-dependent oscillatory shear flow. At the collective level, this results in the development of nematic order in a  suspension of non-interacting particles. For $A\gg 1$, the mean shear flow component is superseded by the oscillatory component, and the particles undergo complex yet synchronized back-and-forth angular oscillations and tumbling motion~(see Sec.~IE-IF of \citep{supplementary_m}). Our results also extend to continuum models of deformable particles by modifying $\beta(r)$, the flow aligning scale~(see Sec.~V of \citep{supplementary_m}). Furthermore, the phenomenon of synchronization-induced nematic order
% observed here from shear oscillations 
occurs in steady shear flow if the non-interacting particles are active. In Sec.~III and IV of \citep{supplementary_m}, we provide mathematical analysis and numerical results to show phase synchronization for particles actively undergo shape oscillations or have periodically varying chiral activity, respectively. Such situations may be relevant for active particles undergoing shape deformations~\citep{PhysRevLett.113.048101,PhysRevLett.114.208101,PhysRevLett.134.038301,PhysRevFluids.4.103302} or in general when the active particles are undergoing periodic modulations in an internal state. Hence, our work opens up new avenues to explore at the intersection of nonlinear dynamics of synchronization and soft \& active matter.

% a dense suspension of nematic particles~\citep{supplementary_m}. For a dense suspense, the equations of spatially-averaged shape parameter and orientation take the same form as Eqs.~\eqref{eq:reqn}-\eqref{eq:thetaeqn} with a rescaling of $\beta(r)$ by $\lambda_0 \beta(r)$ where $\lambda_0>1$ is a flow aligning scale determining the effects of fluid flow on the dense suspension. Since the form of the equations remain same, the same mechanism of phase synchronization is also realized in this system. The nematic order reported in this manuscript via phase-synchronization might have implications for rheological properties in both soft matter systems as well as active matter systems. 

\textit{Acknowledgments.} 
We thank Julia M. Yeomans for useful discussions. I.H. acknowledges funding from the Gould \& Watson Scholarship. S.P.T. thanks the Royal Society and the Wolfson Foundation for the Royal Society Wolfson Fellowship award and acknowledges the support of the Department of Science and Technology, India via the research grant CRG/2023/000169. R.V. acknowledges the support of the Leverhulme Trust [Grant No. LIP-2020-014] and the ERC Advanced Grant ActBio (funded as UKRI Frontier Research Grant EP/Y033981/1). %Add details about Julia's grant ...
%Add a comment about distinguishing our work from droplets and microcapsules in shear flow. In these cases, there is another degree of freedom related to tank treading motion. Here we are only considering two degrees of freedom and their synchronization with external flow.

\bibliography{SYNC_PRL}% Produces the bibliography via BibTeX.

\end{document}